\documentclass[oneside,english]{amsart}
\usepackage{mathptmx}
\usepackage[T1]{fontenc}
\usepackage[latin9]{inputenc}
\usepackage{amsthm}
\usepackage{amstext}
\usepackage{graphicx}
\usepackage{amssymb}
\usepackage[numbers]{natbib}

\makeatletter

\newcommand{\mathcircumflex}[0]{\mbox{\^{}}}

\numberwithin{equation}{section} 
\numberwithin{figure}{section} 

\makeatother

\usepackage{babel}

\begin{document}

\title{Nonassociative quantum theory, emergent probability, and coquasigroup
symmetry}

\author{J.~K\"oplinger}

\address{Jens K\"oplinger: 105 E Avondale Dr, Greensboro NC 27403, USA.}

\email{jens@prisage.com}

\urladdr{http://jenskoeplinger.com/P}

\author{V.~Dzhunushaliev}

\address{Vladimir Dzhunushaliev: Institute for Basic Research, Eurasian National
University, Astana, 010008, Kazakhstan.}

\email{e-mail: vdzhunus@krsu.edu.kg}

\date{21 April 2011}
\begin{abstract}
This paper follows recent steps towards a nonassociative quantum theory
and points out the mathematical structure behind the proposed modifications
to conventional quantum theory. An $N=1$ supersymmetry model and
a strong force glueball ansatz is highlighted. Using nonassociative
complex octonion algebra, it is shown how the Lorentz Lie algebra
can be understood as a four dimensional generalization of the algebra
of spin-1/2 operators in physics. Probability is speculated to become
an emergent phenomenon from some nonassociative geometry in which
to better understand the fluxes involved. A prototype nonassociative
quantum theory in one dimension is brought forward to illustrate how
normed division algebras may aid in modeling isospin properties that
are similar to observed field and particle symmetries in nature. This
prototype is built from a principle of self-duality between types
of active and passive transformations and supplied with a modified
Born rule that models observation, similar to conventional quantum
mechanics. Solutions on the complex numbers, quaternions and octonions
are discussed. The Hopf coquasigroup structure of the octonionic eigenvalue
relation is shown and advertised as a tool for future investigation
into the complete solution set of the model.
\end{abstract}
\maketitle

\section{Introduction}

A primary motivation for work towards a complete, consistent quantum
theory on nonassociative spaces is the desire to model more physical
forces with fewer assumptions. Theoretical reductionism is particularly
important when the consequences of a proposed model are hard or impossible
to measure, as is the case in quantum gravity. How can one unification
proposal be evaluated against another? Today's description of physical
law does not require nonassociativity as a fundamental notion in quantum
mechanics. On the other hand, it is unclear whether quantum gravity
and its unification with the Standard Model may ever be modeled using
conventional associative geometry.

If nonassociativity is a fundamental property of nature, as presumed
here, then it must be shown how today's conventional formulations
may emerge from such foundation without contradicting the observation.
Section \ref{sec:TowardsNonassocQuantTh} follows up on recent investigations
into nonassociative quantum theory and highlights the mathematical
structure of certain modifications to conventional quantum theory:
Nonassociative parts of quantum mechanical operators are unobservable
in principle\citep{Dzhu2007ObsUnobs}, and decompositions exist for
a supersymmetric nonrelativistic Hamiltonian\citep{Dzhun2007NonassocSuperAndHidden},
the spin-$\frac{1}{2}$ operator algebra and Lorentz Lie algebra\citep{Dzhu2009naQFT,Koepl2009octoocto},
and the hypothesized {}``glueball'' particle from strongly interacting
fields\citep{Dzhu2009NonassocDecompStrong,Dzhu2010NonperturbQC,Dzhu2010SU3FluxTube}.
It is shown how probability conservation in conventional quantum mechanics
may become an emergent phenomenon that may be better described in
a nonassociative geometry to be found.

There are many proposals today that introduce nonassociativity into
conventional quantum formulations. More direct approaches may use
nonassociative algebras, such as the octonions or split-octonions,
instead of customary complex number or matrix algebra. More indirect
approaches embed observed Lie group symmetries into the exceptional
Lie groups, which are automorphism groups of types of nonassociative
algebras (for a review see e.g.~\citep{Baez2002TheOctonions}). The
range of envisioned applications spans much of fundamental physics.
A certainly incomplete list of nonassociativity in quantum physics
over the past four decades includes: quark statistics in the Strong
Force\citep{GunayidinGursey1973QuarkStructure,GunaydinGursey1974QuarkStatistics,GunaydinPironRuegg1978OctoQM,Okubo1995IntoOctoNonassocBook},
chirality and triality in fundamental particles\citep{SchrayManogueOcts1994},
Standard Model symmetries from spinors over the division algebras
\citep{Dixon1994DivisionAlgs,Furey2010UnifThIdeal}, the Weak Force
and Yang-Mills instantons\citep{Okubo1995IntoOctoNonassocBook}, octonionic
quantum theory and Dirac equation from left/right-associating operators
\citep{LeoAbdelKhalek1996OctoQM,LeoAbdelKhalek1996OctoDirac,LeoAbdelKhalek1998Octoworld},
fermion generations\citep{DrayManogue1999QuaternionicSpin,DrayManogue1998DimRed},
a geometric relation between Heisenberg uncertainty and the light
cone\citep{Gogb2004OctonionicGeometry}, the Dirac equation with electromagnetic
field\citep{Gogb2005OctoVersionsDiracEqn,Gogb2006OctonionElectrodyn,koepl2007GravEMconicSed},
a four dimensional Euclidean operator quantum gravity\citep{koepl2007hypernumbersRel,Koepl2009octoocto},
Lie group symmetries of the Standard Model \citep{DrayManogue2009OctoSpinorsAndE6,ManogueDray2009OctE6andParticle}
with gravity\citep{Lisi2007E8TOE}, and supersymmetry with the Standard
Model\citep{BaezHuerta2009DivisionAlgsAndSUSY,BaezHuertaDivAlgSUSY2}.
These and other approaches introduce nonassociativity into existing
formulations in physics, which requires modifying some assumptions
while keeping others unchanged. Yet, with all these clues and hints
it is entirely in the open whether a {}``better'' description of
physical law may ever be found this way. If one believes this could
be accomplished, the tantalizing question is: Which, and how many,
of today's paradigms in physics need to be amended?

Section \ref{sec:HopfCoquasiCandidateMethod} proposes a new prototype
nonassociative quantum theory in one dimension that is built from
algebraic and geometric rules. Rather than declaring physical principles
up front (e.g.~conservation of probability, invariance of the speed
of light, equivalence of energy and masses), the model builds wave
functions from self-dual types of transformations. The Born rule which
governs observation in conventional quantum mechanics is modified,
and requires the operator/eigenfunction/eigenvalue relation to be
contained in a complex number subalgebra of the otherwise quaternionic
or octonionic formulation. Solutions exist that are similar to what
one could expect from a physical model.

All work is done under the speculation that the prototype's current
limitation of one spacial dimension may eventually be overcome and
model nature's spacetime as we observe it. One possible way towards
achieving this goal is shown in section \ref{sec:Hopf}. A further
generalized Born rule requires the real eigenvalue of the operator/eigenfunction
relation to remain invariant under changes between equivalent algebras.
An understanding of the complete solution set of such a generalization
appears contingent on a proper mathematical tool. The eigenvalue equations
to be solved are shown to have Hopf coquasigroup structure \citep{KlimMajid2009HopfCoquasigroup}.

\section{Towards a nonassociative quantum theory}

\label{sec:TowardsNonassocQuantTh}Conventional operator quantum mechanics
uses unobservable wave functions $\Psi$ that are decomposed into
orthogonal eigenfunctions $\psi_{n}$ of an operator $H$, to yield
observable eigenvalues $h_{n}$. The expectation value of $H$ over
some configuration space $V$ is then determined through expressions
like:\begin{align*}
\left\langle \Psi\right|H\left|\Psi\right\rangle  & =\sum_{n}h_{n}\int_{V}\psi_{n}^{*}\psi_{n}dV.\end{align*}
Probability density $\rho$ models the relative frequency of occurrence
of measurement outcomes $h_{n}$, and is defined through:\begin{align*}
\rho & :=\left\langle \Psi\right|1\left|\Psi\right\rangle =\sum_{n}\rho_{n}, & \rho_{n} & :=\int_{V}\psi_{n}^{*}\psi_{n}dV.\end{align*}
What is also called the {}``Born rule'' gives the expectation value
of $H$ as:\begin{align}
\left\langle \Psi\right|H\left|\Psi\right\rangle  & =\sum_{n}h_{n}\rho_{n}.\label{eq:expectationValueBornRule}\end{align}
As a special case, the Hamiltonian operator of a quantum mechanical
system allows to describe the time dependency of other operators through
Ehrenfest's theorem. If $H$ is the Hamiltonian and $L$ another operator
on that same system, then the expectation value of $L$ as a function
of time is:\begin{align}
\left\langle \Psi\right|\frac{dL}{dt}\left|\Psi\right\rangle  & =\left\langle \Psi\right|\frac{\partial L}{\partial t}+\frac{\imath}{\hbar}\left[H,L\right]\left|\Psi\right\rangle .\label{eq:defEhrenfestClassical}\end{align}
Here, $\imath$ is the imaginary basis element of the complex numbers,
$b_{\mathbb{C}}=\left\{ 1,\imath\right\} $.

Two operators $H_{1},H_{2}$ with eigenvalues $h_{1n},h_{2n}$ model
physical quantities that can be observed simultaneously only if they
commute:\begin{eqnarray*}
H_{1}\left(H_{2}\left|\Psi\right\rangle \right)=H_{2}\left(H_{1}\left|\Psi\right\rangle \right) & \Longleftrightarrow & h_{1n}\textrm{ and }h_{2n}\textrm{ simultaneously observable}.\end{eqnarray*}
For example, momentum operator $\hat{p}_{i}:=-\imath\hbar\partial/\partial x_{i}$
(with $i=1,2,3$) and angular momentum operator $\hat{L}_{i}:=-\imath\hbar\left(\vec{x}\times\nabla\right)_{i}$
allow only components with same index $i$ to be measured simultaneously
since $\hat{p}_{i}\left(\hat{L}_{i}\left|\Psi\right\rangle \right)=\hat{L}_{i}\left(\hat{p}_{i}\left|\Psi\right\rangle \right)$,
but not two different components since generally $\hat{p}_{i}\left(\hat{L}_{j}\left|\Psi\right\rangle \right)\neq\hat{L}_{j}\left(\hat{p}_{i}\left|\Psi\right\rangle \right)$
for $j\neq i$.

\subsection{Nonassociativity and unobservables}

Noncommutativity of operators from conventional quantum mechanics
is now extended to nonassociativity and speculated to be of use in
a future nonassociative quantum theory. New kinds of operators $Q_{n}$
may in general not satisfy:\begin{align}
\left\langle \Psi\right|\left(Q_{n}\left|\Psi\right\rangle \right) & \neq\left(\left\langle \Psi\right|Q_{n}\right)\left|\Psi\right\rangle .\label{eq:defNonassocOperator}\end{align}
This requires additional rules to be supplied to the Ehrenfest theorem
(\ref{eq:defEhrenfestClassical}) or the Born rule (\ref{eq:expectationValueBornRule}),
to obtain expectation values of the $Q_{n}$, understand their evolution
over time and predict measurement outcomes unambiguously.

This can be realized by having $Q_{n}$ and the $\left|\Psi\right\rangle $
in some nonassociative algebra. Such operators are interpreted to
model {}``unobservables'' that cannot be measured in principle \citep{Dzhu2007ObsUnobs}.
The concept is distinct from conventional {}``hidden variables''
models, which contain information that could in principle be extracted
from the quantum system. An example of an unobservable property in
nature would be the color charge in the Strong Force, a property that
is instrumental in the workings of the force; however, cannot be observed
from the outside. It is pointed out that unobservables do not need
to be quantum contributions on small scales. They may in general be
of the same order of magnitude as conventional observable properties.

There are many ways of bringing this general approach into agreement
with the observation. One way is to decompose known operators into
unobservable parts, define dynamics of these parts and show how the
conventional formulation emerges in the appropriate limit\@. For
example, if $H$ is an operator from conventional quantum mechanics,
it could be made from parts, $H:=Q_{1}Q_{2}$, where the $Q_{1}$
and $Q_{2}$ are unobservables. A nonassociative quantum theory, to
be found, would then have to explain why such decomposition is necessary
or desirable, as opposed to merely being possible.

To give an example, without going into the model assumptions%
\footnote{Here: An ansatz from nonrelativistic $N=1$ supersymmetry.%
}, a Hamiltonian $H$ is proposed in \citep{Dzhun2007NonassocSuperAndHidden,Dzhu2009naQFT}
to be made from $Q_{1}$ and $Q_{2}$ with the following properties:\begin{eqnarray*}
H & := & \frac{1}{2}\left(Q_{1}+Q_{2}\right)^{2},\\
Q_{1}Q_{1}\left|\Psi\right\rangle =Q_{2}Q_{2}\left|\Psi\right\rangle  & = & 0,\\
\left(Q_{1}Q_{2}\right)\left|\Psi\right\rangle  & = & \left(Q_{2}Q_{1}\right)\left|\Psi\right\rangle ,\\
\left\langle \Psi\right|\left(Q_{n}\left|\Psi\right\rangle \right) & \neq & \left(\left\langle \Psi\right|Q_{n}\right)\left|\Psi\right\rangle \qquad\left(n\in\left\{ 1,2\right\} \right),\\
\left\langle \Psi\right|\left(\left(Q_{1}Q_{2}\right)\left|\Psi\right\rangle \right) & = & \left(\left\langle \Psi\right|\left(Q_{1}Q_{2}\right)\right)\left|\Psi\right\rangle .\end{eqnarray*}
The $Q_{1}$ and $Q_{2}$ are unobservables per (\ref{eq:defNonassocOperator}).
These relations can be satisfied when modeling the $Q_{1/2}$ as linear
differential operators and using nonassociative split-octonion algebra
(for details, see \citep{Dzhun2007NonassocSuperAndHidden,Dzhu2009naQFT}).
A new quantum theory could then specify the dynamics of $Q_{1/2}$
and split-octonion wave functions $\left|\Psi\right\rangle $, but
model the observable operator $H=Q_{1}Q_{2}=Q_{2}Q_{1}$ in agreement
with conventional quantum theory.

\subsection{Example: Spin operator and Lorentz Lie algebra from nonassociative
algebra}

For a new nonassociative quantum theory to be useful or desired, it
has to do more than just recreating known physics. There have to be
novel observable effects, or it has to describe known effects using
fewer assumptions. This section gives an example that hints towards
the latter. A nonassociative algebra is shown to have two associative
subalgebras, each of which models an independent effect in physics:
the algebra of spin operators from spin-$\frac{1}{2}$ particles,
and the Lorentz Lie algebra from Special Relativity \citep{Dzhu2008HiddenStructures}.
The finding demonstrates an opportunity for a future quantum theory
that uses nonassociative algebra, to let previously unrelated descriptions
of natural law emerge from a single formalism.

\subsubsection{Algebra of spin-$\frac{1}{2}$ operators}

A spin in physics is a fundamental internal property of particles
or bound quantum systems, such as atomic nuclei. It is independent
from the space-time or energy-momentum parameters that describe other
dynamic properties. A simple example is a spin-$\frac{1}{2}$ particle,
where two spin states are possible when measured along any direction
in space: {}``up'' or {}``down''. Conventional quantum mechanics
describes spin observables through operators $\hat{s}_{j}$:\begin{align*}
\hat{s}_{j} & :=\frac{1}{2}\sigma_{j}\qquad\left(j\in\left\{ 1,2,3\right\} \right).\end{align*}
The index $j$ enumerates three orthogonal spacial axes $x_{j}$ along
which to measure. In the choice of units here%
\footnote{In SI units there is an additional constant here, the Planck constant
$\hbar$. It becomes $1$ in the choice of units in this paper.%
}, only a factor $\frac{1}{2}$ comes with the $\sigma_{j}$, which
are the Pauli matrices over complex numbers. To basis $b_{\mathbb{C}}=\left\{ 1,\imath\right\} $
these are:\begin{align}
\sigma_{1} & :=\left(\begin{array}{rr}
0 & 1\\
1 & 0\end{array}\right), & \sigma_{2} & :=\left(\begin{array}{rr}
0 & -\imath\\
\imath & 0\end{array}\right), & \sigma_{3} & :=\left(\begin{array}{rr}
1 & 0\\
0 & -1\end{array}\right),\label{eq:defPauliMatAlg2}\\
\sigma_{j}\sigma_{j} & =\left(\begin{array}{rr}
1 & 0\\
0 & 1\end{array}\right)=\left(\mathrm{id}\right), & \sigma_{j}\sigma_{k} & =-\frac{\imath}{2}\sum_{l=1}^{3}\epsilon_{jkl}\sigma_{l} &  & \left(j,k\in\left\{ 1,2,3\right\} \right).\nonumber \end{align}
Born's rule gives measurable spin states from eigenfunctions $\left|\Psi\right\rangle $
to the $\sigma_{j}$, so that $\sigma_{j}\left|\Psi\right\rangle =\lambda\left|\Psi\right\rangle $
with real eigenvalues $\lambda=+\frac{1}{2}$ for {}``spin up''
and $\lambda=-\frac{1}{2}$ for {}``spin down'' along an axis of
measurement.

Without addressing physical measurement, the algebra of $\sigma_{j}$
operators (\ref{eq:defPauliMatAlg2}) can be expressed in the associative
complex quaternion algebra. Written to a quaternion basis $b_{\mathbb{H}}=\left\{ 1,i_{1},i_{2},i_{3}\right\} $
and complex number coefficients to $b_{\mathbb{C}}=\left\{ 1,\imath\right\} $,
the $\sigma_{j}$ can be defined as:\begin{align}
\sigma_{j} & :=\imath i_{j}.\label{eq:defPauliMatComplexQuats}\end{align}
On a side note, the imaginary quaternions (here to basis elements
$\left\{ i_{1},i_{2},i_{3}\right\} $) also generate the $\mathfrak{su}\left(2\right)$
Lie algebra.

\subsubsection{Lorentz Lie algebra}

\label{sub:LorentzLieAlg}The Lorentz group is the matrix Lie group
that preserves the quadratic form $\left|\cdot\right|$ on four-vectors
$x:=\left(x_{0},x_{1},x_{2},x_{3}\right)$:\begin{align*}
\left|\cdot\right| & \,:\,\mathbb{R}^{4}\rightarrow\mathbb{R}, & \left|x\right| & :=x_{0}^{2}-x_{1}^{2}-x_{2}^{2}-x_{3}^{2}=x_{0}^{2}-\left\Vert \vec{x}\right\Vert ^{2}.\end{align*}
In Special Relativity in physics, this quadratic form is interpreted
as the metric tensor $\eta$ of Minkowski spacetime:\begin{align*}
\eta_{\mu\nu} & :=\left(\begin{array}{rrrr}
1 & 0 & 0 & 0\\
0 & -1 & 0 & 0\\
0 & 0 & -1 & 0\\
0 & 0 & 0 & -1\end{array}\right), & \left|x\right| & =\sum_{\mu,\nu=0}^{3}x_{\mu}x_{\nu}\eta_{\mu\nu}.\end{align*}
Here, $x_{0}$ is called the {}``time component'' and $\vec{x}:=\left(x_{1},x_{2},x_{3}\right)$
the {}``spacial components'' of the four-vector $x$. Examples for
such four-vectors are energy-momentum $p:=\left(E,p_{1},p_{2},p_{3}\right)=\left(E,\vec{p}\right)$
or space-time intervals $dx:=\left(dt,dx_{1},dx_{2},dx_{3}\right)=\left(dx_{0},d\vec{x}\right)$.
The preserved quadratic form of energy-momentum corresponds to invariant
mass $m^{2}=E^{2}-\left\Vert \vec{p}\right\Vert ^{2}$, and space-time
intervals model invariant proper time $d\tau^{2}=dt^{2}-\left\Vert d\vec{x}\right\Vert ^{2}$.

These physical properties remain unchanged when translating between
equivalent frames of reference. Next to a translation symmetry, the
geometry of Minkowski spacetime is symmetric under rotation in space,
and transformations between uniformly moving, nonaccelerated frames
of reference. The last two symmetries together make up the Lorentz
group. The associated Lie algebra of the generators of Lorentz transformation
$M_{\mu\nu}$ is:\begin{align}
\left[M_{\mu\nu},M_{\rho\sigma}\right] & =\imath\left(\eta_{\nu\rho}M_{\mu\sigma}+\eta_{\mu\sigma}M_{\nu\rho}-\eta_{\mu\rho}M_{\nu\sigma}-\eta_{\nu\sigma}M_{\mu\rho}\right),\label{eq:defLorentzLieAlgebra}\\
x_{\mu}' & :=\sum_{\mu=0}^{3}M_{\mu\nu}x_{\nu},\qquad\mu,\nu,\rho,\sigma\in\left\{ 0,1,2,3\right\} .\nonumber \end{align}
The $M_{\mu\nu}$ rotate the spacial components of a four-vector $x$,
and perform so-called {}``Lorentz boosts''.

Similar to the algebra of spin-$\frac{1}{2}$ operators above, the
Lorentz Lie algebra can be generated with octonions $\mathbb{C}\otimes\mathbb{O}$
to basis $\left\{ 1,\imath\right\} \otimes\left\{ 1,i_{1},\ldots,i_{7}\right\} $
when defining \citep{Dzhu2008HiddenStructures,Koepl2009octoocto}:\begin{align*}
R_{0} & :=\frac{i_{4}}{2}\left(1+\imath\right), & R_{j} & :=\frac{i_{\left(j+4\right)}}{2}\left(1-\imath\right), &  & \left(j\in\left\{ 1,2,3\right\} \right)\\
M_{\mu\nu} & :=\frac{1}{2}\left[R_{\mu},R_{\nu}\right].\end{align*}
All terms of the $M_{\mu\nu}$ are calculated in appendix A, and it
follows directly that the defining relation of the Lorentz Lie algebra
(\ref{eq:defLorentzLieAlgebra}) is satisfied.

\subsubsection{Nonassociative algebra}

The spin-$\frac{1}{2}$ operator generated by $\sigma_{j}$ (\ref{eq:defPauliMatComplexQuats})
and the Lorentz Lie algebra generated by $M_{\mu\nu}$ (\ref{eq:defLorentzLieAlgebra})
are both expressed on complex octonions:\begin{align*}
\sigma_{j} & :=-\frac{\imath}{4}\epsilon_{jkl}R_{k}R_{l}, & M_{\mu\nu} & :=\frac{1}{2}\left[R_{\mu},R_{\nu}\right].\end{align*}

The four $R_{\mu}$ satisfy the additional associator relation:\begin{align*}
\left(R_{\mu},R_{\nu},R_{\rho}\right) & :=\left(R_{\mu}R_{\nu}\right)R_{\rho}-R_{\mu}\left(R_{\nu}R_{\rho}\right)=2\sum_{\sigma,\xi=0}^{3}\epsilon_{\mu\nu\rho\xi}\eta_{\xi\sigma}R_{\sigma}.\end{align*}
One can validate this expression from \begin{align*}
\left(i_{\mu},i_{\nu},i_{\rho}\right) & =2\sum_{\sigma=4}^{7}\varepsilon_{\mu\nu\rho\sigma}i_{\sigma}, & \mu,\nu,\rho & \in\left\{ 4,5,6,7\right\} ,\end{align*}
which is a property of any antiassociative four-tuple in the octonions
(here: $\left\{ i_{4},i_{5},i_{6},i_{7}\right\} $). The Minkowski
tensor $\eta_{\xi\sigma}$ then comes from the difference in sign
in the $\left(1\pm\imath\right)$ factor of $R_{0}$ and the $R_{j}$.

With this, the four element set $\left\{ R_{0},R_{1},R_{2},R_{3}\right\} $
that generates the Lorentz Lie algebra can be viewed as a four dimensional
{}``spacetime'' generalization of the $\left\{ R_{1},R_{2},R_{3}\right\} $
set that generates the algebra of spin-$\frac{1}{2}$ operators, $\sigma_{j}$,
in three dimensional space.

\subsection{Example: Operator algebra of strongly interacting fields and glueball}

The Strong Force in physics is the interaction between building blocks
of matter, the quarks. It is mediated through exchange particles,
the gluons. Both quarks and gluons carry a color charge, but neither
charges nor particles can be isolated or directly observed. This is
known as color confinement in physics.

In the context of this paper, this means that there exist no operators
$A$ in conventional quantum mechanics that would allow measurement
of the color charge with real eigenvalues $a_{n}$ after the Born
rule (\ref{eq:expectationValueBornRule}) to $\left\langle \Psi\right|A\left|\Psi\right\rangle =\sum_{n}a_{n}\rho_{n}$.
This section shows how a certain modification to this rule for observation
in quantum mechanics makes room for nonassociative operator algebras.
This may aid in modeling the \emph{glueball}, a hypothetical particle
that is made from gluons only \citep{Dzhu2010NonperturbQC,Dzhu2010SU3FluxTube}.

\subsubsection{Observables from nonassociative parts of an operator and modified
Born rule}

A product of two operators $A^{B}$ and $A^{C}$ is measured in conventional
quantum mechanics as:\begin{align*}
A^{B}A^{C}\left|\Psi\right\rangle  & \overset{\mathrm{def}}{=}A^{B}\left(A^{C}\left|\Psi\right\rangle \right).\end{align*}
The operators $A^{B},A^{C}$ are now proposed to be made from a product
of operators $e$, $\Phi^{B}$ and $\Phi^{C}$:\begin{align*}
A^{B} & :=e\Phi^{B}, & A^{C} & :=e\Phi^{C}.\end{align*}
The operator algebra $\mathbb{G}$ of the $A^{B}$ and $A^{C}$ is
associative by definition, whereas the $e$, $\Phi^{B}$ and $\Phi^{C}$
are elements in a nonassociative operator algebra%
\footnote{Even though it is not specified what algebras the $\mathbb{A}$ and
$\mathbb{G}$ exactly are, it compares on a very general level to
the $R_{\mu}$ and $\sigma_{j}$ from the previous section. There,
the spin operators $\sigma_{j}$ were elements in the associative
complex quaternion algebra, $\sigma_{j}\in\mathbb{C}\otimes\mathbb{H}$,
whereas the $R_{\mu}$ were from the nonassociative complex octonions,
$R_{\mu}\in\mathbb{C}\otimes\mathbb{O}$. This comparison with the
previous section does not hold much beyond this point, though.%
} $\mathbb{A}$: \begin{align*}
A^{B},A^{C} & \in\mathbb{G}, & e,\Phi^{B},\Phi^{C} & \in\mathbb{A}.\end{align*}

A modification to the Born rule for observability in quantum mechanics
can then be proposed. For operators that are made from nonassociative
parts, measurement requires to reassociate the parts:\begin{align*}
A^{B}A^{C}\left|\Psi\right\rangle  & =A^{B}\left(A^{C}\left|\Psi\right\rangle \right)=\left(e\Phi^{B}\right)\left(\left(e\Phi^{C}\right)\left|\Psi\right\rangle \right)\\
 & =e\left(\Phi^{B}\left(e\left(\Phi^{C}\left|\Psi\right\rangle \right)\right)\right)+m^{BC}\left|\Psi\right\rangle .\end{align*}
The $m^{BC}\left|\Psi\right\rangle $ term is the associator,\begin{align*}
m^{BC} & :=\left(e\Phi^{B}\right)\left(e\Phi^{C}\right)-e\left(\Phi^{B}\left(e\Phi^{C}\right)\right),\\
\left\langle \Psi\right|A^{B}A^{C}\left|\Psi\right\rangle  & :=\left\langle \Psi\right|e\left(\Phi^{B}\left(e\left(\Phi^{C}\left|\Psi\right\rangle \right)\right)\right)+\left\langle \Psi\right|m^{BC}\left|\Psi\right\rangle .\end{align*}
Such operators $A^{B}$ and $A^{C}$ then model an observable physical
quantity only if wave functions $\left|\Psi\right\rangle $ exist
where applying the operators' reassociated constituents yields a real-valued
function $\chi$ and a constant (but not necessarily real) factor
$a^{BC}$as: \begin{align*}
\left\langle \Psi\left(x_{1},x_{2}\right)\right|e\left(x_{1}\right)\left(\Phi^{B}\left(x_{1}\right)\left(e\left(x_{2}\right)\left(\Phi^{C}\left(x_{2}\right)\left|\Psi\left(x_{1},x_{2}\right)\right\rangle \right)\right)\right) & \overset{!}{=}a^{BC}\chi\left(x_{1},x_{2}\right),\end{align*}
\begin{align*}
\Psi & \,:\,\mathbb{R}^{N}\otimes\mathbb{R}^{N}\rightarrow\mathbb{R}^{N}, & x_{1},x_{2} & \in\mathbb{R}^{N},\\
\chi & \,:\,\mathbb{R}^{N}\otimes\mathbb{R}^{N}\rightarrow\mathbb{R}, & a^{BC} & =\mathrm{const}.\end{align*}

\subsubsection{Glueball}

When the $A^{B}$ and $A^{C}$ operators are interpreted as unobservable
Strong Force fields and charges, the methodology can be applied to
a quantum system that interacts purely through fields. This is possible
in principle when particles that mediate a force carry charge themselves,
as is the case with the gluons in the strong force. A bound state
between gluons only, without quarks, has been referred to as \emph{glueball}
in the literature. But conventional treatment of the strong force
currently indicates that such a particle either doesn't exist, or
if it exists it would always be in a superposition with regular particles
from bound quarks, where it would be indistinguishable therefrom.

A solution for the glueball has been brought forward \citep{Dzhu2010NonperturbQC,Dzhu2010SU3FluxTube},
by adapting an approach similar to this section and constraining degrees
of freedom to model a force with known characteristics from the strong
force in physics. Since the new glueball solution is obtained from
a quantum theory that generally does not reduce to conventional quantum
theory, there is an opportunity to predict novel effects from this
nonstandard treatment.

\subsection{Emergent probability from a nonassociative geometry?}

Conventional quantum mechanics specifies a conservation rule for probability
density $\rho$ and flux $\vec{j}$:\begin{align*}
\frac{\partial}{\partial t}\rho+\mathrm{div}\,\vec{j} & =0.\end{align*}
This relation can be extended to more than three spacial dimensions
in the $\mathrm{div}\,\vec{j}$ term when the underlying geometry
is locally flat and differentiable. For a single time axis and three
or more spacial dimensions, an $n$-dimensional vector space over
the reals $\mathbb{R}^{n}$ can be equipped with a quadratic metric
of the form:\begin{align*}
ds^{2} & :=dt^{2}-\sum_{j=1}^{n-1}dx_{j}^{2}.\end{align*}
For a static volume $X$ with no flux $\vec{j}$ on the surface, probability
density $\rho$ is then required to be conserved as a function of
time:\begin{align*}
\frac{\partial}{\partial t}\int_{X}\rho\, d^{n-1}x & =0.\end{align*}

When allowing nonassociative wave functions $\left|\Psi\right\rangle $
where generally $\left\langle \Psi\right|\left(Q\left|\Psi\right\rangle \right)\neq\left(\left\langle \Psi\right|Q\right)\left|\Psi\right\rangle $,
this requires additional conditions on how to extract observable values
$h_{n}$ with probabilities $\rho_{n}$. Recalling the Born rule for
observability, clarification is required at the fundamental level
of quantum mechanics:

\begin{align*}
\left\langle \Psi\right|Q\left|\Psi\right\rangle  & \overset{?}{=}\sum_{n}h_{n}\int_{V}\psi_{n}^{*}\psi_{n}dV, & \rho_{n} & \overset{?}{=}\int_{V}\psi_{n}^{*}\psi_{n}dV.\end{align*}
Rather than trying to somehow fit nonassociative algebra into these
relations from conventional quantum mechanics, it is now speculated
for classical probability to become an emergent phenomenon, where
nonassociative values of $\left|\Psi\right\rangle $ suggest some
kind of nonassociative geometry in which to better understand the
fluxes involved. This notion of theoretical reductionism ultimately
has to prove itself in an actual model. It needs to confirm or not
whether simplification is indeed achievable, and describe natural
law with fewer assumptions. It is noted that probability doesn't have
to be abandoned as a concept altogether. But there may be an opportunity
for modeling physical law in approaches where nonconservation of probability
forced investigators to abort in the past.

\section{Prototype nonassociative quantum theory in one dimension}

\label{sec:HopfCoquasiCandidateMethod}A new nonassociative quantum
theory will have to be consistent in itself and reproduce known results
in parameter ranges that have been tested experimentally. To be considered
for comparison with existing theories, it will have to predict new
testable effects or describe known effects with fewer assumptions.

This section brings forward a prototype for such a theory that is
built from algebraic rules: There are types transformations in a vector
space, a self-duality principle, and an eigenvalue invariance condition.
In strong simplification as compared to nature, all physical fields
and charges are placed along a single, preferred real axis in $\mathbb{R}^{d}$
. Wave functions $\psi$ are made from two types of transformations,
active $T^{A}$ and passive $T^{P}$, which map the preferred real
axis into the unit sphere in $d$ dimensions, $S^{d-1}$. Active and
passive transformations are considered dual to one another, and relate
through a condition that can be satisfied in the normed division algebras
$\mathbb{C}$ ($d=2$), $\mathbb{H}$ ($d=4$), and $\mathbb{O}$
($d=8$). Fields and particles in physics are mapped to active and
passive transformations respectively. The mathematical duality between
the allowed types of transformations becomes a self-duality principle
between physical fields and particles. Physical measurement requires
an eigenfunction/eigenvalue rule similar to the Born rule in conventional
quantum mechanics, with the additional requirement that the eigenvalue
relation must be reducible to a complex number description%
\footnote{This requirement leads to a class of quaternion and octonion algebras
that are equivalent in the sense that the eigenvalue relation in complex
number form remains unchanged when switching between equivalent algebras.
This is discussed in section \ref{sec:Hopf}.%
}.

Solutions are shown and the prototype is advertised for further exploration.
Using complex numbers and asking for the influence of many fields
on a single particle, the solutions are the Dirac equation with $1/r$
fields if a timeless physical world would only have one dimension
in space. Quaternionic solutions exist only if all fields (or particles)
are local to the point that marks a particle in the complex numbers.
There must be at least two contributing fields%
\footnote{Or particles; fields and particles are required to be equivalent duals.
The side note {}``(or particles)'' is therefore omitted going forward
when talking about fields, but it is always implied.%
} that cannot be observed or probed independently. The real eigenvalue
from the modified Born rule remains invariant under general rotation
of the imaginary quaternion basis. Therefore, the eigenvalue relation
is said to have local $\mathrm{SU}\left(2\right)$ Lie group symmetry.
Octonionic solutions are further restricted by requiring at least
three contributing fields. One solution is shown and said to have
local $\mathrm{G}_{2}$ symmetry. Without claiming completeness, the
solution set of the prototype appears wide enough to sufficiently
resemble physical reality, given the model's current limitation of
only one spacial dimension and no time concept.

\subsection{Configuration space, self-duality, active and passive transformations}

The model is built in a $d$-dimensional vector space over the reals,
$\mathbb{R}^{d}$. One preferred axis in $\mathbb{R}^{d}$ corresponds
to physical space and is denoted with $x$. A physical system is described
through a combination of \emph{active} and \emph{passive} transformations,
which map the $x$-axis onto the unit sphere in $\mathbb{R}^{d}$:\begin{align*}
T^{A},T^{P} & \,:\,\mathbb{R}\rightarrow S^{d-1}.\end{align*}
Active and passive transformations are modeled as exponentials%
\footnote{A different type of active and passive transformation was discussed
in \citep{Gogber2008SplitOctoRotations}, where a {}``passive''
transformation was a rotation of the coordinate basis elements of
$\mathbb{R}^{8}$ that leaves the norm of a split-octonion invariant.
{}``Active'' rotations were actions on the split-octonion basis
elements that result in another split-octonion basis.%
} using normed division algebras:\begin{eqnarray*}
\theta & \in & \begin{cases}
\mathbb{C} & \left(d=2\right),\\
\mathbb{H} & \left(d=4\right),\\
\mathbb{O} & \left(d=8\right),\end{cases}\qquad\left|\theta\right|=1,\qquad\theta^{2}=-1,\\
T^{A}\left(x\right) & := & \left|x-a\right|^{\theta t_{A}}:=\exp\left(\theta t_{A}\ln\left|x-a\right|\right)\in S^{d-1},\\
T^{P}\left(x\right) & := & \theta^{\left(x-a\right)t_{P}}:=\exp\left(\left(x-a\right)t_{P}\ln\theta\right)\\
 & = & \exp\left(\theta t_{P}\left(x-a\right)\left(\frac{\pi}{2}+2\pi N\right)\right)\in S^{d-1},\\
x,a,t_{A},t_{P} & \in & \mathbb{R},\, x\neq a,\, N\in\mathbb{Z}.\end{eqnarray*}
The natural logarithm with real argument, $\ln\left|x-a\right|$,
is chosen to be real-valued by definition. This choice omits possible
terms $\pm2\pi\theta$ in the exponent of $T^{A}\left(x\right)$.
For $x\neq a$ it is always possible to find $\left\{ x,a,t_{A},t_{P},\theta\right\} $
such that:\begin{eqnarray*}
T^{A}\left(x\right) & \overset{!}{=} & T^{P}\left(x\right),\\
\exp\left(\theta t_{A}\ln\left|x-a\right|\right) & \overset{!}{=} & \exp\left(\theta t_{P}\left(x-a\right)\left(\frac{\pi}{2}+2\pi N\right)\right),\\
\Rightarrow\, t_{A} & = & t_{P}\frac{x-a}{\ln\left|x-a\right|}\left(\frac{\pi}{2}+2\pi N\right)\qquad\textrm{for }x\neq a.\end{eqnarray*}
This relation allows to call $T^{A}$ and $T^{P}$ equivalent duals
under a map $\widetilde{\cdot}$ that exchanges base and exponent:\begin{align*}
T^{A} & \sim\alpha^{\beta}, & T^{P} & \sim\beta^{\alpha}, & \alpha^{\beta} & \sim\widetilde{\beta^{\alpha}}.\end{align*}

\subsection{Fields and particles}

When modeling the electromagnetic force in physics, the {}``first
quantization'' particle point of view of Quantum Electrodynamics
is fully equivalent to the {}``second quantization'' field point
of view of Quantum Field Theory. The speculation here is that this
equivalence can be carried forward for modeling physical forces beyond
electromagnetism, given a suitable quantum theory (to be found). The
mathematical duality between $T^{A}$ and $T^{P}$ becomes a \emph{self-duality
principle} when declaring active transformations to model physical
fields, and passive transformations to model physical particles. The
proposed field-particle duality is thereby reflected in mathematical
properties of the model. It is arbitrary to assign active transformations
$T^{A}$ to fields, as opposed to particles. Since $T^{A}$ and $T^{P}$
are equivalent duals, the choice is irrelevant for the model predictions
as long as one adheres to it throughout the entire calculation.

Interaction between many particles and fields is then modeled by an
effective transformation $T^{\mathrm{eff}}$ that is a product of
any number of active and passive transformations:\begin{eqnarray*}
f & : & T_{1}\otimes\ldots\otimes T_{n}\rightarrow T^{\mathrm{eff}},\\
T_{i} & \in & \left\{ T_{i}^{A},T_{i}^{P}\right\} \quad\textrm{for }i=1,\ldots,n,\\
T^{\mathrm{eff}} & : & \mathbb{R}\rightarrow S^{d-1}.\end{eqnarray*}
The tensor symbol $\otimes$ indicates possible complex number, quaternion,
and octonion multiplication rules. Terms from the $\left\{ T_{i}\right\} $
may be expressed as their Taylor polynomials. After choosing an algebra,
the $f$ become polynomial functions in $\mathbb{R}^{d}$. The effective
transformations $T^{\mathrm{eff}}$ will also be written with the
symbol $\psi$ to be similar to notation customary for wave functions
in physics.

\subsection{Physical measurement, modified Born rule, and select solutions}

The Born rule from conventional quantum mechanics governs physical
measurement. As was recalled in section \ref{sec:TowardsNonassocQuantTh},
operators $\hat{D}$ model observable physical quantities when they
act on wave functions $\psi\equiv T^{\mathrm{eff}}$ that are eigenfunctions
to $\hat{D}$ with real eigenvalues $\lambda$. An additional condition
is now supplied for the prototype quantum theory here, which requires
the eigenfunctions $\psi$ to fall into a complex number subspace
of the algebra:\begin{align*}
\hat{D}\psi & \overset{!}{=}\lambda\psi, & \lambda & \in\mathbb{R}, & \psi & \in\mathbb{C}\subset\left\{ \mathbb{H},\mathbb{O}\right\} .\end{align*}

\subsubsection{Complex numbers}

In the complex number case to basis $b_{\mathbb{C}}:=\left\{ 1,\imath\right\} $,
a solution exists for a linear differential operator $\hat{D}$, wave
function $\psi$ and eigenvalue $m\in\mathbb{R}$:\begin{eqnarray*}
\hat{D} & := & -\imath\frac{\partial}{\partial x}-\sum_{i=1}^{n-1}\frac{t_{i}}{x-a_{i}},\\
\psi & := & \imath^{t_{n}x}\left(\prod_{i=1}^{n-1}\left|x-a_{i}\right|^{\imath t_{i}}\right)\\
 & = & \exp\left(\imath\pi t_{n}x\left(\frac{1}{2}\pm2M\right)\right)\prod_{i=1}^{n-1}\exp\left(\imath t_{i}\ln\left|x-a_{i}\right|\right),\end{eqnarray*}
\begin{align}
\Rightarrow\,\hat{D}\psi & =m\psi, & m & =\pi t_{n}\left(\frac{1}{2}\pm2M\right), & M & \in\mathbb{N}.\label{eq:ComplexesOpEqDirac}\end{align}
The wave function $\psi$ contains a product of active transformations
$T_{i}^{A}$ which are interpreted as fields or external influences
on a quantum system:\begin{align*}
T_{i}^{A} & =\left|x-a_{i}\right|^{\imath t_{i}}.\end{align*}
The passive transformation $T^{P}$ is interpreted as particle property,
or characteristic property of the system under investigation:\begin{align*}
T^{P} & =\imath^{t_{n}x}.\end{align*}

In comparison with physics, the Dirac equation with electromagnetic
field can be written with complex-valued $4\times4$ matrices $\gamma_{j}$,
four vectors $\Psi$ and spacetime coordinates $\left\{ x_{0},\ldots,x_{3}\right\} $
as:\begin{align*}
\hat{D}_{\mathrm{EM}} & :=\sum_{j=0}^{3}\gamma_{j}\left(-\imath\frac{\partial}{\partial x_{j}}-\sum_{i=1}^{n-1}\frac{t_{i}}{x_{j}-a_{i}}\right),\\
\gamma_{j} & \in\mathbb{C}^{4}\times\mathbb{C}^{4},\qquad\Psi\in\mathbb{C}^{4},\\
\hat{D}_{\mathrm{EM}}\Psi & =m\Psi.\end{align*}
These equations use four spacetime coordinates $x_{j}$ instead of
a single coordinate $x$ in (\ref{eq:ComplexesOpEqDirac}), and four
$\gamma_{j}$ matrices instead of a mere multiplicative identity.
The operator equation (\ref{eq:ComplexesOpEqDirac}) is interpreted
as a state equation of a test particle under the influence of linear
superpositioned $1/x$-type fields of different strength $t_{i}$
and poles at places $a_{i}$ along a single $x$ axis. The particle
is located at $x=0$ and has a characteristic property $m\in\mathbb{R}$.
On this primitive level it supports the conjecture that the prototype
quantum theory is similar enough to known physics to be of interest
for further investigation of spacetime-internal isospin properties
of a quantum system. Of course, the argument cannot be made conclusively
as long as it is unknown how, or even if, today's description of observed
spacetime can be made to emerge in a future generalization of the
prototype.

Equation (\ref{eq:ComplexesOpEqDirac}) models a test particle under
the influence of $\left(n-1\right)$ fields. Any amount of further
external influences can be supplied independently as:\begin{align*}
\psi' & :=\psi\exp\left(\imath\alpha\right), & \alpha & \in\mathbb{R}.\end{align*}
Using verbiage customary in physics, this property is now called a
\emph{global symmetry}. Additional fields can be supplied independently
at any place along the real axis through superposition with the existing
fields. Since the $\exp\left(\imath\alpha\right)$ terms generate
the $\mathrm{U}\left(1\right)$ Lie group under multiplication, \begin{align*}
\left\{ \exp\left(\imath\alpha\right),\,\alpha\in\mathbb{R}\right\}  & \cong\mathrm{U}\left(1\right),\end{align*}
the eigenvalue relation is said to have global $\mathrm{U}\left(1\right)$
symmetry.

\subsubsection{Quaternions}

Not every combination of active and passive transformations in the
quaternions can be written as a single effective transformation using
complex numbers only. For example, a combination of a single active
and passive transformation may fall into a complex number subspace
only if both are already contained in that same subspace:\begin{align*}
T_{1}^{A} & :=\left|x-a_{1}\right|^{\theta_{1}t_{1}}=\exp\left(\theta_{1}t_{1}\ln\left|x-a_{1}\right|\right),\\
T_{2}^{P} & :=\theta_{2}^{t_{2}\left(x-a_{2}\right)}=\exp\left(\theta_{2}\pi\left(x-a_{2}\right)t_{2}\left(\frac{1}{2}+2M_{2}\right)\right),\\
\psi & :=T_{1}^{A}T_{2}^{P},\qquad\psi\in\mathbb{C}\,\Longleftrightarrow\,\theta_{1}=\pm\theta_{2}.\end{align*}
Due to the different $x$-dependency in $T_{1}^{A}$ and $T_{2}^{P}$,
the two unit quaternions $\theta_{1}$ and $\theta_{2}$ must necessarily
be linear dependent for $\psi$ to remain in the same complex number
subalgebra for any $x\neq a_{1},a_{2}$.

A truly quaternionic solution may therefore only come from wave functions
that are a product of a single type of transformation, active or passive.
The following examines the example of interacting particles%
\footnote{The same reasoning from the example is valid for interacting fields.
This must be the case since the transformations satisfy the self-duality
requirement.%
}. A pair of passive transformations in the quaternions is in general:\begin{align*}
T_{1}^{P}T_{2}^{P} & =\exp\left(\theta_{1}\pi\left(x-a_{1}\right)t_{1}\left(\frac{1}{2}+2M_{1}\right)\right)\exp\left(\theta_{2}\pi\left(x-a_{2}\right)t_{2}\left(\frac{1}{2}+2M_{2}\right)\right)\\
 & =\exp\left(\theta_{1}c_{1}\right)\exp\left(\theta_{1}d_{1}x\right)\exp\left(\theta_{2}c_{2}\right)\exp\left(\theta_{2}d_{2}x\right),\end{align*}
\begin{align*}
c_{1} & :=-\pi a_{1}t_{1}\left(\frac{1}{2}+2M_{1}\right), & d_{1} & :=\pi t_{1}\left(\frac{1}{2}+2M_{1}\right),\\
c_{2} & :=-\pi a_{2}t_{2}\left(\frac{1}{2}+2M_{2}\right), & d_{2} & :=\pi t_{2}\left(\frac{1}{2}+2M_{2}\right).\end{align*}
The $c_{1,2},d_{1,2}\in\mathbb{R}$ and $M_{1,2}\in\mathbb{Z}$ are
constants and independent of $x$. The imaginary unit quaternions
$\theta_{1,2}$ are elements of the $\mathfrak{su}\left(2\right)$
Lie algebra made from the imaginary quaternion basis elements $\left\{ i_{1},i_{2},i_{3}\right\} $.
The Baker-Campbell-Hausdorff formula for Lie algebras gives existence
of an imaginary unit quaternion $\tilde{\theta}$ from the same $\mathfrak{su}\left(2\right)$
algebra such that:\begin{align*}
\exp\left(\tilde{\theta}\tilde{d}\right) & =\exp\left(\theta_{1}d_{1}\right)\exp\left(\theta_{2}d_{2}\right).\end{align*}
$\tilde{d}$ is a real constant. The $x$-dependency can be written
to a single unit quaternion $\tilde{\theta}$:\begin{align*}
T_{1}^{P}T_{2}^{P} & =\exp\left(\theta_{1}c_{1}\right)\exp\left(\theta_{1}d_{1}x\right)\exp\left(\theta_{2}d_{2}x\right)\exp\left(\theta_{2}c_{2}\right)\\
 & =\exp\left(\theta_{1}c_{1}\right)\exp\left(\tilde{\theta}\tilde{d}x\right)\exp\left(\theta_{2}c_{2}\right).\end{align*}

To express $\psi$ using a single term $\exp\left(\tilde{\theta}\tilde{d}x\right)$
there may be two cases:
\begin{enumerate}
\item The $\theta_{1},\theta_{2}$ are identical (except for a possible
sign change). In this case, $\psi$ is fully contained in a complex
number subalgebra within the quaternions and the eigenvalue equation
reduces to the complex number case.
\item The $a_{1}$ and $a_{2}$ are the same, $a:=a_{1}=a_{2}$, so that
all particles are located at the same position along $x$. This makes
$c_{1}$ and $c_{2}$ multiples of $d_{1}$ and $d_{2}$ by the same
real factor, $c_{1,2}=ad_{1,2}$, and allows for a new quaternionic
solution $\exp\left(\tilde{\theta}\tilde{d}\left(x-a\right)\right)=\exp\left(\theta_{1}d_{1}\left(x-a\right)\right)\exp\left(\theta_{2}d_{2}\left(x-a\right)\right)$. 
\end{enumerate}
In the second case there is a new quaternionic type of solution where
$\psi$ can be expressed as:\begin{align*}
\psi & =\exp\left(\tilde{\theta}\tilde{d}\left(x-a\right)\right).\end{align*}
An operator $\hat{D}$ exists that commutes with $\psi$ and has a
real eigenvalue $\tilde{d}$:\begin{align*}
\hat{D} & :=-\tilde{\theta}\frac{\partial}{\partial x}, & \hat{D}\psi & =\psi\overleftarrow{\hat{D}}=\tilde{d}\psi.\end{align*}
This satisfies the modified Born rule as the eigenvalue relation can
be reduced to a complex number subalgebra.

It may become complicated to find an explicit expression for $\tilde{\theta}$
and $\tilde{d}$ from a given $\left\{ d_{1},d_{2},\theta_{1},\theta_{2}\right\} $.
But with its existence proven it is concluded that the prototype quantum
theory yields a novel type of solution for the quaternion case. All
particles have to be at the same place $x=a$ which makes it a \emph{local}
solution. Finding a $\tilde{\theta}$ depends on both $\theta_{1}$
and $\theta_{2}$. It is not possible anymore to supply a third particle
with arbitrary $\theta_{3}$ independently, since any new term with
a $\theta_{3}$ generally requires a change of $\tilde{\theta}$.
This is different from the notion of describing influences from a
given system on an independent test particle, as it was possible in
the complex number case.

The $\left\{ \theta_{1},\theta_{2},\tilde{\theta}\right\} $ in the
quaternion example of two interacting particles are elements of an
$\mathfrak{su}\left(2\right)$ Lie algebra. The example can be extended
to any number of interaction partners since the Baker-Campbell-Hausdorff
formula may be repeated any number of times, as long as one remains
within the same Lie algebra. The quaternionic solution is therefore
said to have \emph{local} $\mathrm{SU}\left(2\right)$ \emph{symmetry}.
It models an internal \emph{isospin} property that does not depend
on $x$. This resembles observed properties from the Weak Force in
physics.

\subsubsection{Octonions}

\label{sub:octonionPrototype}Use of octonion algebra justifies calling
the formulation here a \emph{nonassociative} prototype quantum theory.
Wave functions $\psi$ are made from transformations $T_{i}$ which
contain a number of octonionic imaginary unit vectors $\left\{ \theta_{i}\right\} $.
These fall into one of four cases:
\begin{enumerate}
\item All $\theta_{i}$ are identical (except for a possible difference
in sign) and the eigenvalue relation reduces to the complex number
case.
\item All $\theta_{i}$ are associative under multiplication and are therefore
elements of the same $\mathfrak{su}\left(2\right)$ algebra. This
reduces to the quaternion case.
\item The automorphism group of the $\left\{ \theta_{i}\right\} $ is $\mathrm{G}_{2}$,
which is the automorphism group of the octonions.
\item The automorphism group of the $\left\{ \theta_{i}\right\} $ is $\mathrm{SU}\left(3\right)$,
which is the subgroup of $\mathrm{G}_{2}$ that leaves one imaginary
octonion unit unchanged.
\end{enumerate}
Cases 3 and 4 are new with the octonions. They require a product of
at least three $T_{i}$, since any two octonion basis elements are
always part of an $\mathfrak{su}\left(2\right)$ algebra and therefore
have an automorphism group no larger than $\mathrm{SU}\left(2\right)$.

Octonions $\left\{ \theta_{i}\right\} $ generally don't form a Lie
algebra. It is not possible to find an octonion $\tilde{\theta}$
and real number $\tilde{d}$ for any given $\theta_{i},d_{i}$ ($i=1,2,3$)
such that:\begin{eqnarray*}
\left(\exp\left(\theta_{1}d_{1}\right)\exp\left(\theta_{2}d_{2}\right)\right)\exp\left(\theta_{3}d_{3}\right) & \overset{?}{=} & \exp\left(\tilde{\theta}\tilde{d}\right).\end{eqnarray*}
But there are subalgebras in the octonions that are larger than the
quaternions, for which the Baker-Campbell-Hausdorff equation is still
applicable. The nonassociative Lie algebras $\mathfrak{g}_{2}$ or
$\mathfrak{su}\left(3\right)\subset\mathfrak{g}_{2}$ can be expressed
in terms of octonions (e.g.~\citep{Dixon1994DivisionAlgs}). The
$\mathfrak{g}_{2}$ can be written as algebra of derivations over
the octonions, $\mathfrak{der}\left(\mathbb{O}\right)$, in form of
\citep{Baez2002TheOctonions,Schafer1995nonassIntro}:\begin{align*}
D_{u,v}\left(a\right) & =\left[\left[u,v\right],a\right]-3\left(\left(uv\right)a-u\left(va\right)\right), & u,v,a & \in\mathbb{O}.\end{align*}
Since the $D_{u,v}\left(a\right)$ form a Lie algebra it is possible
to find $\tilde{\theta}$ and $\tilde{d}$ for any given $\theta_{i},d_{i}$
($i=1,2,3$) and $u,v$:\begin{align}
\left(\exp\left(D_{u,v}\left(\theta_{1}\right)d_{1}\right)\exp\left(D_{u,v}\left(\theta_{2}\right)d_{2}\right)\right)\exp\left(D_{u,v}\left(\theta_{3}\right)d_{3}\right) & =\exp\left(D_{u,v}\left(\tilde{\theta}\right)\tilde{d}\right).\label{eq:waveFunctionAsG2}\end{align}
A wave function $\psi$ and operator $\hat{D}$ exist that model a
solution for the nonassociative prototype quantum theory:

\begin{align*}
\psi & :=\exp\left(\tilde{\theta}_{D}\tilde{d}\left(x-a\right)\right), & \hat{D} & :=-\tilde{\theta}_{D}\frac{\partial}{\partial x}, & \hat{D}\psi & =\tilde{d}\psi.\end{align*}

Solutions in this ansatz are similarly restricted as in the quaternion
case. All particles%
\footnote{The same reasoning applies to fields as well and is always implied,
just as in the quaternion case. A wave function $\psi_{F}:=\exp\left(\tilde{\theta}_{D}\tilde{d}\ln\left(x-a\right)\right)$
would be an eigenfunction to $\hat{D}_{F}:=-\tilde{\theta}_{D}\left(x-a\right)\left(\partial/\partial x\right)$
with eigenvalue $\lambda=\tilde{d}$.%
} have to be located at the same place. Observable wave functions are
made from interactions between particles, but without the ability
of inserting an independent test probe. For a truly octonionic eigenvalue
relation at least three particles are required to model a wave function
(\ref{eq:waveFunctionAsG2}). Through the $\left\{ D_{u,v}\left(\theta_{i}\right)\right\} $,
their imaginary unit vectors $\left\{ \theta_{i}\right\} $ may generate
the nonassociative algebra $\mathfrak{g}_{2}$ or its $\mathfrak{su}\left(3\right)$
subalgebra.

On this very high level it appears that the prototype continues to
be a candidate for future usefulness in physics. The octonion case
contains spaces that have the observed $\mathrm{SU}\left(3\right)$
symmetry of the Strong Force between quarks. The minimum number of
quarks that can enter into a bound state is three, not counting quark-antiquark
states%
\footnote{Time as a concept is absent from the current prototype theory, and
since particle-antiparticle duality is related to time symmetry in
nature, it is conjectured that the prototype quantum theory does not
contradict this.%
}.

\section{Hopf coquasigroup symmetry of the octonionic eigenvalue relation}

\label{sec:Hopf}This section is looking at ways to extend the prototype
nonassociative quantum theory to more than one dimension in space,
as nature is obviously more than one dimensional. An equivalence class
for normed division algebras is introduced, and a further generalized
Born rule for physical observation requires a real eigenvalue to remain
invariant under changes between equivalent algebras. The prototype
nonassociative quantum theory in one dimension from the previous section
is shown to be contained in such an approach, which now leaves room
for supplying additional dimensions independently. Understanding the
mathematical structure of the new equivalence class is advertised
as key to understanding the physical meaning of its associated solution
spaces.

\subsection{A further generalization to the Born rule}

It appears natural to assume that a solution along the $x$ axis,\begin{align}
\hat{D}_{x}\psi\left(x\right) & =\tilde{d}\psi\left(x\right),\label{eq:eigenvalueWithSingleAxis}\end{align}
should still be valid if properties $\phi\left(y\right)$ along some
other axis $y$ orthogonal to $x$ exist:\begin{eqnarray}
\hat{D}_{x}\psi\left(x\right)=\tilde{d}\psi\left(x\right) & \Longrightarrow & \hat{D}_{x}\psi\left(x\right)\phi\left(y\right)=\tilde{d}\psi\left(x\right)\phi\left(y\right).\label{eq:eigenvalueWithOtherPart}\end{eqnarray}
Even though the factors may be using nonassociative octonion algebra,
it is not needed to set brackets on the right-hand side since $\hat{D}_{x}$
only acts on $\psi\left(x\right)$, $\hat{D}_{x}\psi\left(x\right)$
is in the same complex number subalgebra as $\psi\left(x\right)$,
and all normed division algebras are alternative.

The requirement to express the entire eigenvalue equation (\ref{eq:eigenvalueWithOtherPart})
in the complex plane appears too narrow for a modified Born rule,
as it would restrict the $\phi\left(y\right)$ to the same complex
number subalgebra for any $y$, without need. To make room for expansion,
a class is now introduced such that the real eigenvalue in (\ref{eq:eigenvalueWithOtherPart})
is to remain invariant when switching between equivalent normed division
algebras. Any two algebras are equivalent under this class if they
share the same axes in their associative imaginary basis triplets,
but allow for a change in sign (or order, or parity) of these triplets.
Equivalent algebras will be labeled $N=0,\ldots,M-1$.

Parity of imaginary quaternion basis triplets gives rise to algebraic
noncommutativity. A quaternion algebra $\mathbb{H}\left[0\right]$
may be defined with $i_{1}i_{2}:=-i_{2}i_{1}:=i_{3}$, and another
$\mathbb{H}\left[1\right]$ with $i_{1}i_{2}:=-i_{2}i_{1}:=-i_{3}$.
These $M=2$ algebras are equivalent under the new class. In the octonions,
parity of its basis triples gives rise to noncommutativity as well
as nonassociativity. There are $M=16$ equivalent octonion algebras.

If the eigenvalue equation (\ref{eq:eigenvalueWithSingleAxis}) is
confined to a complex number subalgebra, then $\tilde{d}$ remains
invariant under changes between equivalent algebras. The one dimensional
prototype quantum theory is therefore contained in an extension that
requires eigenvalue invariance under changes of algebra in its generalized
Born rule. In turn this allows for introduction of independent axes
as in (\ref{eq:eigenvalueWithOtherPart}) that may be modeled in other
algebra subspaces.

In symbolic form, wave functions $\psi$ are polynomial functions
$\psi\equiv f\left[N\right]$ made from polynomials $f$ supplied
with an algebra multiplication that is octonionic in general. A functor
$A$ maps the polynomial $f\in P$ into the set of polynomial functions
$\left\{ f\left[N\right]\right\} $ that are made from equivalent
algebras:\begin{align*}
A & \,:\, P\rightarrow\left\{ \mathbb{R}\otimes\ldots\otimes\mathbb{R}\rightarrow S^{7}\right\} , & A\left(f\right) & :=\left\{ f\left[N\right],\, N=0,\ldots,M-1\right\} .\end{align*}
Here, the variable list of real parameters $\mathbb{R}\otimes\ldots\otimes\mathbb{R}$
denote the possible physical axes or dimensions. The 7-sphere $S^{7}$
is the unit sphere in $\mathbb{R}^{8}$. Requiring an eigenvalue $\lambda\in\mathbb{R}$
to remain invariant under changes of algebra then becomes the generalized
Born rule for observation: \begin{align}
a & \overset{!}{=}\lambda f\left[0\right]\textrm{ for any }a\in A\left(\hat{D}f\right);\qquad\textrm{or equivalently:}\nonumber \\
A\left(\hat{D}f\right) & \overset{!}{=}\lambda\left\{ \underbrace{f\left[0\right],\ldots,f\left[0\right]}_{M\textrm{ times}}\right\} .\label{eq:defGenerlizedBornRuleDef}\end{align}
The index $\left[0\right]$ is by choice and refers to one of the
$M$ possible algebras. Any of the equivalent algebras from $A$ may
be selected at will for a certain index number. Once selected, however,
this choice has to remain throughout the entire calculation.

\subsection{Hopf quasigroup structure}

In the octonions to basis $b_{\mathbb{O}}=\left\{ 1,i_{1},\ldots,i_{7}\right\} $
there are seven associative permutation triplets of imaginary basis
elements, which are now chosen for an octonion algebra $\mathbb{O}\left[0\right]$
(i.e., $\mathbb{O}$ with index $\left[0\right]$) as:\begin{align}
\mathbb{O}\left[0\right] & :=\left\langle \mathbb{R}^{8},+,\times\right\rangle ,\nonumber \\
i_{\mu}\times i_{\nu} & :=\epsilon_{\mu\nu\rho}i_{\rho}-\delta_{\mu\nu}\textrm{, with}\nonumber \\
\mu\nu\rho & \in t_{\mathbb{O}\left[0\right]}:=\left\{ 123,761,572,653,145,246,347\right\} ,\label{eq:chosenOzeroTriplets}\\
1\times i_{\mu} & :=i_{\mu}\times1=i_{\mu}.\nonumber \end{align}
The product of any three imaginary basis elements is nonassociative
when these basis elements are not contained in a single permutation
triplet%
\footnote{Other choices of triplet labeling are of course possible, for example,
the cyclically symmetric $\left\{ 124,235,346,457,561,672,713\right\} $
used by Dixon \citep{Dixon1994DivisionAlgs}. The choice here has
$\left\{ i_{1},i_{2},i_{3}\right\} $ recalling the associative triplet
from the quaternions and $\left\{ i_{4},i_{5},i_{6},i_{7}\right\} $
as a nonassociative quadruplet that extends quaternions to the octonions.%
} from $t_{\mathbb{O}\left[0\right]}$ (\ref{eq:chosenOzeroTriplets}).
It follows directly from the $\epsilon_{\mu\nu\rho}$ that even permutations
of $\left\{ \mu\nu\rho\right\} $ produce the identical algebra, whereas
odd permutations change the sign of the corresponding product of basis
elements. An odd permutation can be understood as changing the parity
of the triplet.

There are seven basis element triplets in $t_{\mathbb{O}\left[0\right]}$
which allow for $2^{7}$ possible combinations of sign changes. However,
only $16$ of the combinations generate an alternative composition
algebra that is octonion. These $M=16$ combinations are written as
$\left\{ t_{\mathbb{O}\left[N\right]}\textrm{ with }N=0,\ldots,15\right\} $
and represent the set of equivalent algebras $\left\{ \mathbb{O}\left[N\right]\right\} $.

To construct these, one can start from a given $\mathbb{O}\left[0\right]$
and four duality automorphisms $\mathcal{T}_{0},\ldots,\mathcal{T}_{3}$
that act on the $\left\{ \mathbb{O}\left[N\right]\right\} $:\begin{align*}
\mathcal{T}_{n} & :\left\{ \mathbb{O}\left[N\right]\right\} \rightarrow\left\{ \mathbb{O}\left[N\right]\right\} , & \left\{ \left(\mathrm{id}\right),\mathcal{T}_{n}\right\}  & \cong\mathbb{Z}_{2},\\
\mathcal{T}_{n}\mathcal{T}_{n} & =\left(\mathrm{id}\right), & n & \in\left\{ 0,1,2,3\right\} .\end{align*}
$\mathbb{Z}_{2}$ is the cyclic group with two elements. When acting
on the $t_{\mathbb{O}\left[N\right]}$ the $\mathcal{T}_{n}$ either
leave the parity of a permutation triplet unchanged, $\left(\mathrm{id}\right)$,
or swap it, $\left(\mathrm{sw}\right)$:\begin{align}
\mathcal{T}_{0} & :=\left\{ \left(\mathrm{id}\right),\left(\mathrm{id}\right),\left(\mathrm{id}\right),\left(\mathrm{id}\right),\left(\mathrm{sw}\right),\left(\mathrm{sw}\right),\left(\mathrm{sw}\right)\right\} ,\label{eq:defOctoAutomorphisms}\\
\mathcal{T}_{1} & :=\left\{ \left(\mathrm{sw}\right),\left(\mathrm{sw}\right),\left(\mathrm{sw}\right),\left(\mathrm{sw}\right),\left(\mathrm{id}\right),\left(\mathrm{id}\right),\left(\mathrm{id}\right)\right\} ,\nonumber \\
\mathcal{T}_{2} & :=\left\{ \left(\mathrm{id}\right),\left(\mathrm{sw}\right),\left(\mathrm{id}\right),\left(\mathrm{sw}\right),\left(\mathrm{sw}\right),\left(\mathrm{id}\right),\left(\mathrm{sw}\right)\right\} ,\nonumber \\
\mathcal{T}_{3} & :=\left\{ \left(\mathrm{id}\right),\left(\mathrm{id}\right),\left(\mathrm{sw}\right),\left(\mathrm{sw}\right),\left(\mathrm{id}\right),\left(\mathrm{sw}\right),\left(\mathrm{sw}\right)\right\} .\nonumber \end{align}
All possible combinations of the $\left\{ \mathcal{T}_{n}\right\} $
acting on $t_{\mathbb{O}\left[0\right]}$ then generate the $16$
triplet sets $t_{\mathbb{O}\left[N\right]}$ for the $\mathbb{O}\left[N\right]$
respectively. For previous approaches that use this construction see
e.g. the {}``left-handed'' and {}``right-handed'' multiplication
tables from \citep{Lockyer2008Octospace}, or the group action $T$
from \citep{SchrayManogueOcts1994} (equation 30 therein). Octonions
that are here mapped through $\mathcal{T}_{0}$ are called {}``opposite
algebra'' in \citep{SchrayManogueOcts1994} (equation 33 therein)
and correspond to octonionic spinors of opposite chirality. Whereas
$\mathcal{T}_{0}$ changes the parity of three triplets, the $\left\{ \mathcal{T}_{1},\mathcal{T}_{2},\mathcal{T}_{3}\right\} $
each change the parity of four triplets. $\mathcal{T}_{0}$ is an
algebra isomorphism that transitions between opposite algebras of
different chirality \citep{SchrayManogueOcts1994}. It is not an isomorphism
in the sense that opposite algebras could be transformed into one
another through transformation of the basis vectors in $\mathbb{R}^{8}$
alone \citep{Lockyer2008Octospace} (they cannot). The combined $\mathcal{T}_{0}\mathcal{T}_{1}$
inverts the sign of all seven nonreal octonion elements and corresponds
to complex conjugation.

\begin{figure}
\centering{}\caption{\label{fig:T1T2T3Fano-1}All unique automorphisms from repeat application
of the $\left\{ \mathcal{T}_{1},\mathcal{T}_{2},\mathcal{T}_{3}\right\} $
can be graphed in the Fano plane (left), where the product of each
two automorphisms on a line yields the third. Together with identity
$\left(\mathrm{id}\right)$ this forms the group $\mathbb{Z}_{2}^{3}=\mathbb{Z}_{2}\times\mathbb{Z}_{2}\times\mathbb{Z}_{2}$
(right).}
\includegraphics[clip,scale=0.7]{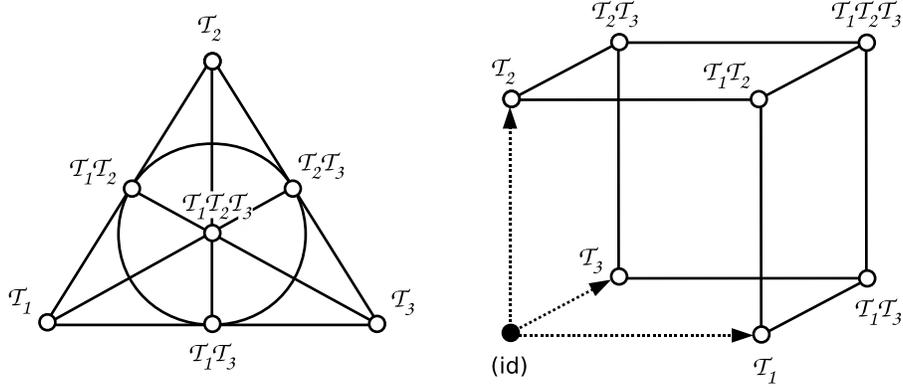}
\end{figure}

The structure of the generalized Born rule on octonions (\ref{eq:defGenerlizedBornRuleDef})
is therefore given by the structure of the $\mathcal{T}_{n}$ from
(\ref{eq:defOctoAutomorphisms}). For a select $n$, the pair $\left\{ \left(\mathrm{id}\right),\mathcal{T}_{n}\right\} $
forms the two element cyclic group $\mathbb{Z}_{2}$. The possible
unique combinations of the $\left\{ \mathcal{T}_{1},\mathcal{T}_{2},\mathcal{T}_{3}\right\} $
form the set\[
\left\{ \mathcal{T}_{1},\,\mathcal{T}_{2},\,\mathcal{T}_{3},\,\mathcal{T}_{1}\mathcal{T}_{2},\,\mathcal{T}_{1}\mathcal{T}_{3},\,\mathcal{T}_{2}\mathcal{T}_{3},\,\mathcal{T}_{1}\mathcal{T}_{2}\mathcal{T}_{3}\right\} \]
which transitions between octonions $\mathbb{O}\left[N\right]$ of
the same chirality. It can be graphed in the Fano plane where the
combination of any two automorphisms on a line yields the third one
(figure \ref{fig:T1T2T3Fano-1}).%
 Together with the identity element, $\left(\mathrm{id}\right)$,
this forms the group%
\footnote{The structure of octonion algebra and its relation to $\mathbb{Z}_{2}^{3}$
and Hadamard transforms is also investigated in \citep{AlbuMajidQuasialg}.%
} $\mathbb{Z}_{2}^{3}=\mathbb{Z}_{2}\times\mathbb{Z}_{2}\times\mathbb{Z}_{2}$.

Together with the chirality-changing $\left\{ \left(\mathrm{id}\right),\mathcal{T}_{0}\right\} \cong\mathbb{Z}_{2}$,
the automorphism group between all the $f\left[N\right]$ is then:\begin{align*}
\mathrm{Aut}\left(\left\{ f_{\mathbb{O}}\left[N\right]\right\} \right) & \cong\mathbb{Z}_{2}^{4}.\end{align*}
The $f\left[N\right]$ themselves are functions that map their of
real arguments into the 7-sphere $S^{7}$, the unit sphere in $\mathbb{R}^{8}$:\begin{eqnarray*}
f\left[N\right] & : & \mathbb{R}\otimes\ldots\otimes\mathbb{R}\rightarrow S^{7}.\end{eqnarray*}
Writing $\left[S^{7}\right]$ as the set of all functions on the 7-sphere,
the symmetry $S^{D}$ of the generalized Born rule (\ref{eq:defGenerlizedBornRuleDef})
becomes:\begin{align*}
S^{D} & \cong\left[S^{7}\right]\rtimes\mathbb{Z}_{2}^{4}.\end{align*}
Such $S^{D}$ is not a group due to nonassociativity of the $\left[S^{7}\right]$.
It may instead have Hopf (co)quasigroup structure as in \citep{KlimMajid2009HopfCoquasigroup}.
If true, it can be equipped with a differential calculus and Fourier
transformation \citep{Klim2010IntegralHopfQuasi}. This mathematical
flexibility would make it a promising structure to investigate the
solution set of the generalized Born rule, which in turn would allow
to identify applicability in physics.

\subsection{Next steps and outlook}

The prototype nonassociative quantum theory in one dimension was developed
as a self-consistent formalism under the speculation that it may be
further developed into a working quantum theory for the description
of nature. Time and space will need to be modeled, and the set of
solutions needs to be understood much deeper. Only then will it be
possible to compare it with other models that are built from observed
or speculated properties of nature.

Active and passive transformations here use exponentiation between
an imaginary vector of unit length and a real number. These are the
central pieces for modeling physical fields and particles. As a morphism
over a two dimensional vector space, exponentiation $\mathcircumflex\,:\,\mathbb{R}^{2}\otimes\mathbb{R}^{2}\rightarrow\mathbb{R}^{2}$
in the complex numbers $\mathbb{C}$ is noncommutative, nonassociative,
nonalternative, a left-inverse is generally different from a right
inverse, and the morphism doesn't distribute over addition. One might
speculate about building new kinds of algebras from requiring existence
of an exponential function that preserves a certain geometric simplicity,
rather than attempting to preserve algebraic rules from pairwise morphisms
(commutativity, associativity, distributivity, and similar). Two such
examples in the two dimensional plane have been brought forward \citep{ShusterKoeplWSpace,ShusterKoeplPQSpace},
and may be of interest for evaluating new kinds of transformations
for applicability in modeling nature.

In all, mathematical properties from nonassociative algebras, spinors,
symmetries, and observed properties of nature at the smallest scales
continue to offer enigmatic similarity, yet it is unclear whether
nonassociativity in physics may ever overcome its current status of
being an incidental curiosity: Are known circumstantial evidence and
suspicious parallels the results of a yet undiscovered theory of consequence?
Time (emergent or not) will tell.

\section*{Acknowledgments}

\thanks{Many thanks to the conference organizers of the {}``2nd Mile High
Conference on Nonassociative Mathematics'' at Denver University,
CO (2009), as well as the {}``Special Session on Quasigroups, Loops,
and Nonassociative Division Algebras'' at the AMS Fall Central Section
Meeting at Notre Dame University, South Bend, IN (2010), to allow
presentation of material from this paper. We are grateful for the
NSF travel grant that allowed VD to participate in person in Denver.
Our best thanks extend to Tevian Dray, Shahn Majid, John Huerta, and
Geoffrey Dixon for open discussions, criticism, and thoughts that
helped develop the material. A special thank you to the referee of
the initial version of the paper, for going to extraordinary length
and detail in the review.}

\section*{Appendix A: Lorentz Lie algebra from nonassociative product}

\label{sec:AppALorentzLieAlg}\setcounter{equation}{0}\renewcommand{\theequation}{A.\arabic{equation}} Section
\ref{sub:LorentzLieAlg} shows the Lorentz Lie algebra,\begin{align}
\left[M_{\mu\nu},M_{\rho\sigma}\right] & =\imath\left(\eta_{\nu\rho}M_{\mu\sigma}+\eta_{\mu\sigma}M_{\nu\rho}-\eta_{\mu\rho}M_{\nu\sigma}-\eta_{\nu\sigma}M_{\mu\rho}\right),\label{eq:LLA2}\\
 & \qquad\textrm{with }\mu,\nu,\rho,\sigma\in\left\{ 0,1,2,3\right\} ,\nonumber \end{align}
and states that this relation can be satisfied in the algebra of complex
octonions $\mathbb{C}\otimes\mathbb{O}$ to basis $\left\{ 1,\imath\right\} \otimes\left\{ 1,i_{1},\ldots,i_{7}\right\} $
when defining \citep{Dzhu2008HiddenStructures,Koepl2009octoocto}:\begin{align*}
R_{0} & :=\frac{i_{4}}{2}\left(1+\imath\right), & R_{j} & :=\frac{i_{\left(j+4\right)}}{2}\left(1-\imath\right) &  & \left(j\in\left\{ 1,2,3\right\} \right),\\
M_{\mu\nu} & :=\frac{1}{2}\left[R_{\mu},R_{\nu}\right].\end{align*}
This appendix calculates relation (\ref{eq:LLA2}) explicitly from
the $R_{\mu}$ to provide proof:

Written in matrix form, the $M_{\mu\nu}$ are:\begin{align*}
M_{\mu\nu}=\frac{1}{2}\left[R_{\mu},R_{\nu}\right] & =\frac{1}{2}\left(\begin{array}{rrrr}
0 & i_{1} & i_{2} & i_{3}\\
-i_{1} & 0 & \imath i_{3} & \,-\imath i_{2}\\
-i_{2} & \,-\imath i_{3} & 0 & \imath i_{1}\\
-i_{3} & \imath i_{2} & -\imath i_{1} & 0\end{array}\right).\end{align*}
The possible combinations of indices $\left\{ \mu,\nu,\rho,\sigma\right\} $
from the Lorentz Lie algebra (\ref{eq:LLA2}) fall in the following
four cases:
\begin{itemize}
\item The case $M_{\mu\nu}=M_{\rho\sigma}$ is trivially satisfied.
\item If $\mu=\nu$ or $\rho=\sigma$ then either $M_{\mu\nu}=0$ or $M_{\rho\sigma}=0$.
The four terms of the right-hand side of relation (\ref{eq:LLA2})
cancel each other out pairwise.
\item If all four elements in $\left\{ \mu,\nu,\rho,\sigma\right\} $ are
different then $M_{\mu\nu}$ must be $\pm\imath M_{\rho\sigma}$:\begin{align*}
M_{01} & =\frac{1}{2}\left[R_{0},R_{1}\right]=\frac{i_{1}}{2}=-\imath\frac{1}{2}\left[R_{2},R_{3}\right]=-\imath M_{23},\\
M_{02} & =\frac{1}{2}\left[R_{0},R_{2}\right]=\frac{i_{2}}{2}=\imath\frac{1}{2}\left[R_{1},R_{3}\right]=\imath M_{13},\\
M_{03} & =\frac{1}{2}\left[R_{0},R_{3}\right]=\frac{i_{3}}{2}=-\imath\frac{1}{2}\left[R_{1},R_{2}\right]=-\imath M_{12}.\end{align*}
This makes the commutator $\left[M_{\mu\nu},M_{\rho\sigma}\right]=0$
as required per (\ref{eq:LLA2}).
\item The remaining case $\mu=\rho$ and $\nu\neq\sigma$ yields:\begin{align*}
\left[M_{01},M_{02}\right] & =\frac{i_{3}}{2}=-\imath\eta_{00}M_{12}, & \left[M_{01},M_{03}\right] & =-\frac{i_{2}}{2}=-\imath\eta_{00}M_{13},\\
\left[M_{12},M_{13}\right] & =-\frac{i_{1}}{2}=-\imath\eta_{11}M_{23}, & \left[M_{10},M_{13}\right] & =\frac{\imath i_{3}}{2}=-\imath\eta_{11}M_{03},\\
\left[M_{20},M_{21}\right] & =\frac{\imath i_{1}}{2}=-\imath\eta_{22}M_{01}, & \left[M_{20},M_{23}\right] & =\frac{\imath i_{3}}{2}=-\imath\eta_{22}M_{03},\\
\left[M_{30},M_{31}\right] & =\frac{\imath i_{1}}{2}=-\imath\eta_{33}M_{01}, & \left[M_{30},M_{32}\right] & =\frac{\imath i_{2}}{2}=-\imath\eta_{33}M_{02}.\end{align*}
All other index combinations are obtained from either switching the
arguments of the commutator bracket, or from switching the indices
of both $M$ terms. From the Minkowski tensor $\eta$ in (\ref{eq:LLA2})
there will only be one nonzero term on the right-hand side. These
are exactly the terms obtained from the complex octonion algebra of
the $R_{\mu}$ above. $\square$\end{itemize}

\end{document}